\RequirePackage{fix-cm}
\documentclass[twocolumn]{svjour3}          
\smartqed  
\usepackage{graphicx}
\usepackage{nicefrac}
\usepackage{hyperref}
\usepackage{color}
\usepackage{amsmath}
\usepackage{amssymb}
\usepackage{eucal}

\newcommand{\Tt}[1]{\mathbf{#1}}

\usepackage{booktabs}

\usepackage{subfigure}

\begin{document}

\title{Numerical simulation of skin transport using Parareal
\thanks{This research is funded by the Deutsche Forschungsgemeinschaft (DFG) as part of the "Exasolvers" project in the Priority Programme 1648 ''Software for Exascale Computing'' (SPPEXA) and by the Swiss National Science Foundation (SNSF) under the lead agency agreement as grant SNF-145271. The research of A.K., D.R., and R.K. is also funded through the FUtuRe SwIss Electrical InfraStructure (FURIES) project of the Swiss Competence Centers for Energy Research (SCCER).}
}


\author{Andreas Kreienbuehl \and
            Arne Naegel \and
            Daniel Ruprecht \and
            Robert Speck\and
            Gabriel Wittum\and
            Rolf Krause
}

\authorrunning{Kreienbuehl, Naegel, Ruprecht, Speck, Wittum, Krause} 

\institute{A. Kreienbuehl \and D. Ruprecht \and R. Krause \at
Institute of Computational Science, Faculty of Informatics, Universit\`a della Svizzera italiana, Via Giuseppe Buffi 13, CH-6904 Lugano, Switzerland	
           \and
           Arne Naegel \and Gabriel Wittum \at
           Goethe-Center for Scientific Computing, Goethe-University Frankfurt, Kettenhofweg 139, 60325 Frankfurt a.M., Germany
           \and
           Robert Speck \at
	  Forschungszentrum J\"ulich GmbH, Institute for Advanced Simulation, J\"ulich Supercomputing Centre, Wilhelm-Johnen-Strasse, DE-52425 J\"ulich, Germany
}

\date{Received: date / Accepted: date}

\maketitle

\begin{abstract}
\textit{In-silico} investigation of skin permeation is an important but also computationally demanding problem.
To resolve all scales involved in full detail will not only require exascale computing capacities but also suitable parallel algorithms.
This article investigates the applicability of the time-parallel Parareal algorithm to a brick and mortar setup, a precursory problem to skin permeation.
The C{\tt++} library \texttt{Lib4PrM} implementing Parareal is combined with the \texttt{UG4} simulation framework, which provides the spatial discretization and parallelization. 
The combination's performance is studied with respect to convergence and speedup.
It is confirmed that anisotropies in the domain and jumps in diffusion coefficients only have a minor impact on Parareal's convergence.
The influence of load imbalances in time due to differences in number of iterations required by the spatial solver as well as spatio-temporal weak scaling is discussed.
\keywords{Skin transport \and Parareal \and space-time parallelism \and weak scaling \and load balancing}
\end{abstract}

\section{Introduction}
Permeation of chemical substances through human skin is an interesting and important process e.g. for the development of cosmetics or drugs. \textit{In-vitro} studies with humans constitute the ``gold standard'' but they are expensive and limited by ethical and practical concerns.
Here, \textit{in-silico} studies are a viable alternative. They have been successfully used in the past (cf. the reviews in \cite{ADDRSkin2013,Querleux2014}) and can be expected to become even more important in the future. They allow for hypothesis testing and may lead to experiments through which effects not known today could be discovered.

Yet, numerical simulations in this field are demanding in terms of computational resources. The problem covers vastly different physical scales and, in case a complex full-fledged model is used, massive computational parallelism needs to be exploited to reach reasonable times-to-solutions. Therefore, many interesting aspects such as substructures of lipid bilayers or networks of keratin filaments \cite{Wang2006,Wang2007} have not yet been investigated in three spatial dimensions (3D) using numerical simulations. Finally, modern imaging techniques make resolving a spectral range of a few nanometers possible, which results in ``big data'' for analyses. Understanding the functional mechanism of the skin is thus a candidate from the life sciences for applying exascale computing.

Considering the technology trend towards more and more parallelism, the application of new parallel methods to the problem investigated here becomes relevant.
Promising candidates for such methods are parallel-in-time integration methods that can add a direction of concurrency in addition to spatial parallelization, e.g., as used here, parallel multi-grid.
In recent years, time-parallel methods have matured from a mainly mathematical concept to an approach with demonstrated efficiency in massively parallel computations \cite{SpeckEtAl2012,RuprechtEtAl2013_SC}. 
They have been listed as a direction for mathematical research with the potential to help reaching exascale computing~\cite{DongarraEtAl2014}.

One of the most widely investigated parallel-in-time methods is Parareal, introduced in 2001 by Lions, Maday and Turinici \cite{LionsEtAl2001}.
It has been used for benchmark problems motivated by applications from fields as diverse as plasma physics \cite{SamaddarEtAl2010}, computational fluid dynamics \cite{Celledoni2009,Randles2014} or quantum chemistry \cite{BylaskaEtAl2013}.
Improvements with respect to implementation are considered e.g. in \cite{Aubanel2011,ElwasifEtAl2011}.
Parareal's most appreciated aspect is probably that it is non-intrusive and rather easy to integrate into existing codes.
Its drawback, on the other hand, is a quite severe bound on achievable parallel efficiency.
However, because several other ``across-the-step'' time-parallel methods share similar features with Parareal (e.g. PITA \cite{FarhatEtAl2003}, MGRIT \cite{FalgoutEtAl2014_MGRIT} or PFASST \cite{EmmettMinion2012}), studying Parareal's performance often already gives important insights.

Theoretical estimates for stability and convergence of Parareal for linear diffusive problems with constant coefficients can be found in \cite{GanderVandewalle2007}.
Theory for diffusive problems with constant coefficients can also be found in \cite{Bal2005}.
For 2D diffusion with space-time dependent coefficients, numerical experiments showed only a marginal reduction in convergence speed \cite{RuprechtEtAl2015_DDM}.
In \cite{ArteagaEtAl2015}, the small impact of a time dependent viscosity on Parareal's convergence for an advection-diffusion problem is demonstrated.
\textit{However, performance for a 3D diffusive problem on a complex geometry with anisotropies has not yet been studied.}

In preparation for the eventual application of Para\-real to skin permeation, this article provides an investigation of Parareal's performance for a 3D brick and mortar problem. 
From our point of view, this model serves as an excellent benchmark because it features challenges resulting from a complex anisotropic geometry and from jumping coefficients, which need to be resolved adaptively over long time intervals. 
Although locally the mathematical formulation of the brick and mortar problem is clear, the global picture is highly complex and linked to a real world application requiring a sound simulation infrastructure in terms of numerical methods and software.
Here, we employ the simulation framework \texttt{UG4} \cite{VogelUG2013} for the spatial discretization and linear solvers, for which excellent parallel scaling has been demonstrated~\cite{ReiterUG2013}. We parallelize \texttt{UG4}'s serial temporal solvers through the C{\tt++} Parareal library \texttt{Lib4PrM}, which is integrated as a plugin.

The present article establishes the principle applicability of Parareal to the 3D brick and mortar problem. In doing so, it identifies a set of relevant issues to be tackled in order to develop an improved parallel-in-time integrator that can deliver reasonable efficiency for the skin transport problem.
In particular, load balancing in time is identified as a critical issue when combining implicit methods for a complex PDE with Parareal. Because the number of iterations of the spatial solver varies in time, balancing temporal subintervals in Parareal simply by the number of time steps induces load imbalances which can affect speedup.

\section{Problem and methods}\label{sec:methods}
\begin{figure*}[th]
\centering
    \subfigure[with cornoecytes (yellow)]{\includegraphics[width=0.4\textwidth]{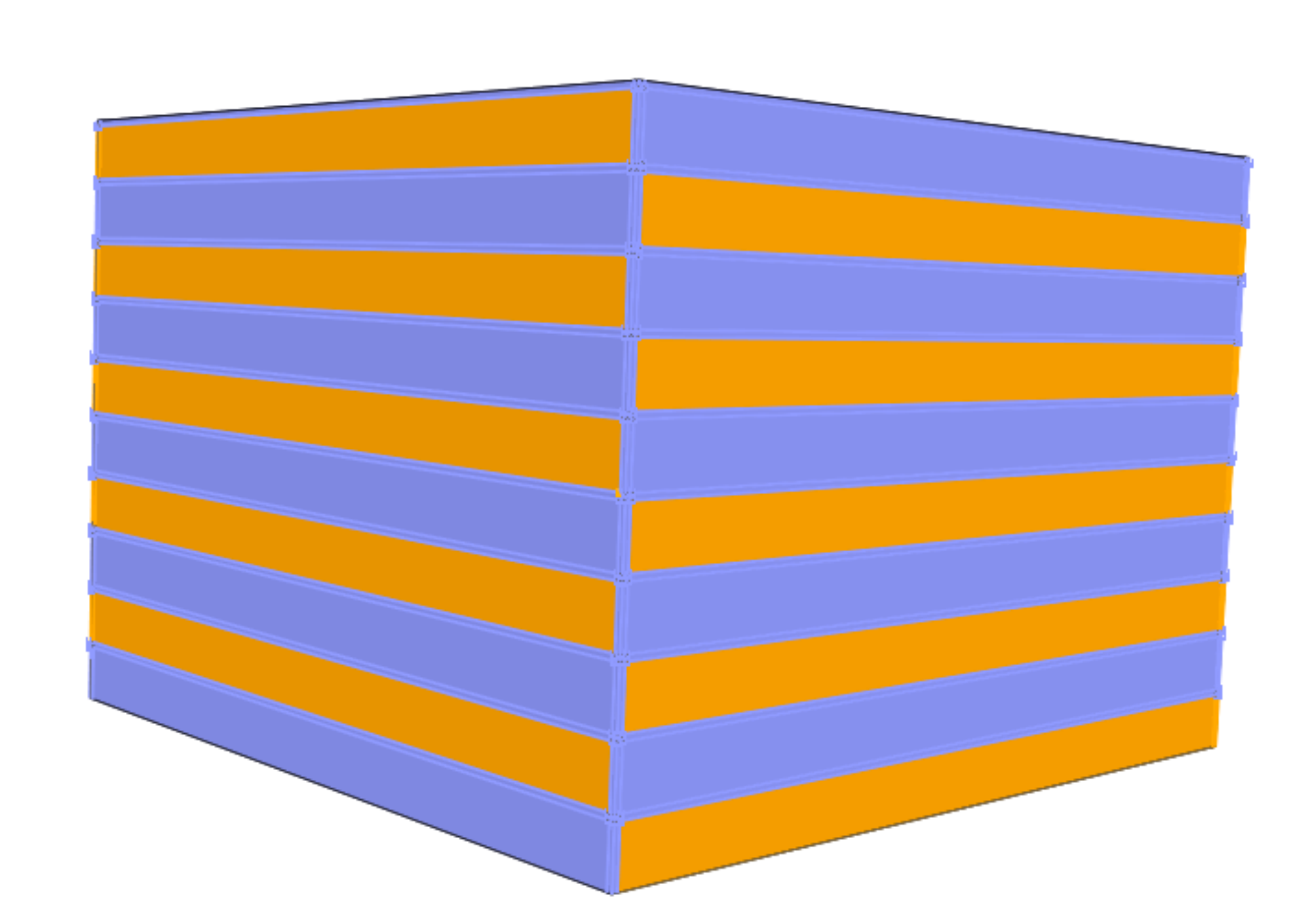}}
    \subfigure[corneocytes removed]{\includegraphics[width=0.4\textwidth]{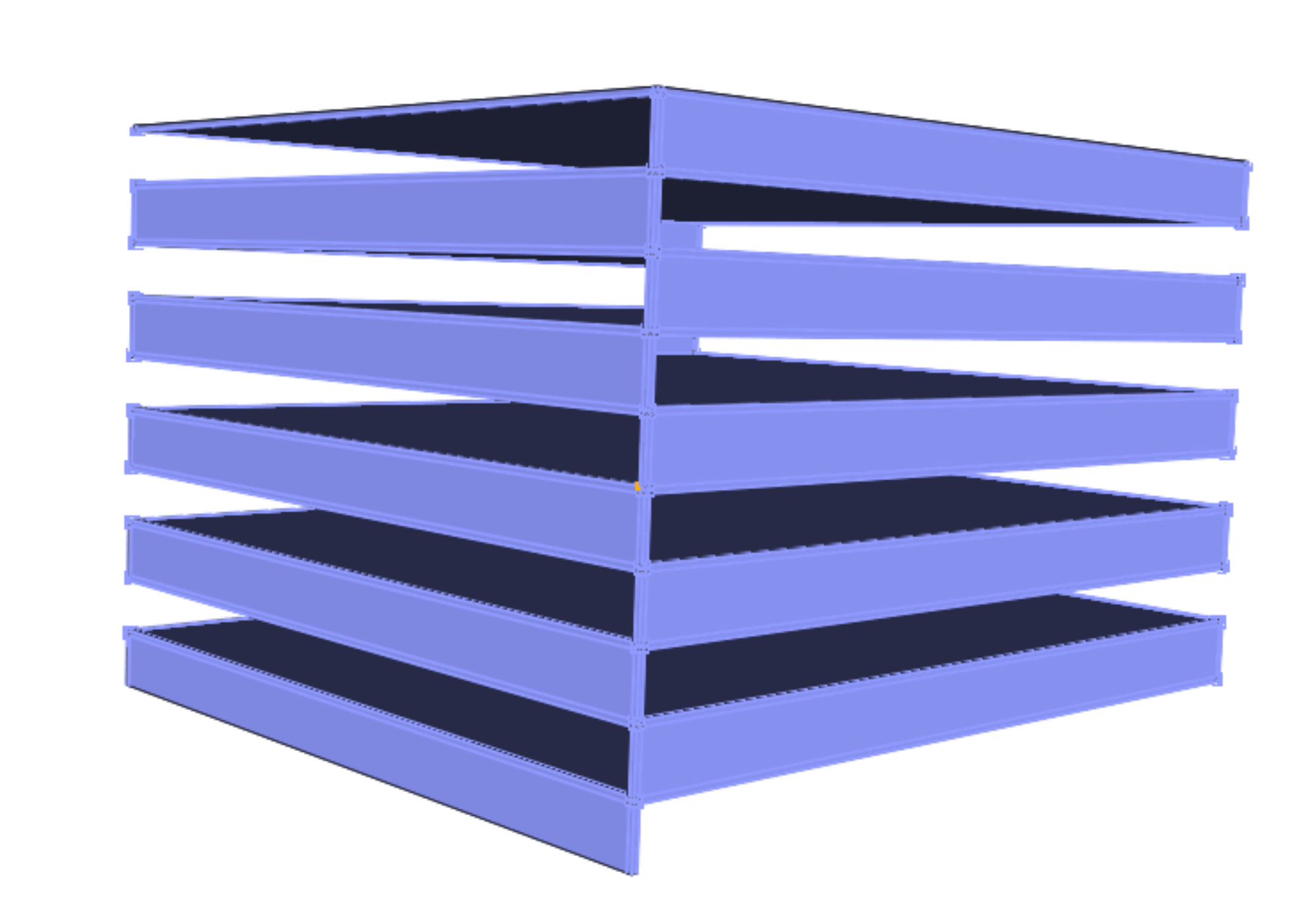}}
    \caption{A sketch of the 3D brick and mortar problem is shown\label{fig:brick}. The  geometry has ten layers of corneocytes ($\Omega_{\text{cor}}$, yellow) that are embedded in a matrix of lipid bilayers ($\Omega_{\text{lip}}$, blue).}
\end{figure*}
As a test case, we study a simplified version of the 3D brick and mortar problem introduced earlier in \cite{Rim2008,Naegel2009}. 
The benchmark is defined on a biphasic domain $\Omega\subset\mathbb{R}^3$ that  consists of two disjoint subsets $\Omega_{\text{cor}}, \Omega_{\text{lip}}\subset \mathbb{R}^3$ representing the so called corneocyte and lipid phase respectively. To be more specific, $\Omega$ is the interior of the union $\overline \Omega_{\text{cor}} \cup \overline \Omega_{\text{lip}}$ of closures. On this geometry we solve a diffusion equation with space-dependent diffusion coefficients. The evolution of the drug concentration $c_{\text{p}}(\Tt{x},t)$, $\text{p}\in\{\text{cor,lip}\}$, is modeled by the equation
\begin{equation}
	\partial_t c_{\text{p}}(\Tt{x},t) = \nabla \cdot \left(  D_{\text{p}} ( \Tt{x}) \nabla c_{\text{p}}(\Tt{x},t)\right)
\end{equation}
with $\Tt{x} \in \Omega_{\text{p}}$, $t \in [0,T]$, and a phase-dependent diffusion coefficient
\begin{equation}
	D_{\text{p}}(\Tt{x})
	=\left\{
		\begin{array}{c @{ \ : \ }l}
			D_{\rm cor} & \Tt{x} \in \Omega_{\text{cor}}, \\
			D_{\rm lip} & \Tt{x} \in \Omega_{\text{lip}}. 
		\end{array}
	\right.
\end{equation}
For the simulation time we use $T=\frac{\lambda^2}{6 D_{\text{eff}}}$ which is the characteristic \emph{lag time} of the problem. It is defined in terms of the membrane thickness $\lambda$ and the effective (homogenized) diffusion coefficient $D_{\text{eff}}$~\cite{Naegel2009}.
In terms of boundary conditions we have an interior phase boundary $\Gamma = \overline \Omega_{\text{cor}} \cap \overline \Omega_{\text{lip}}$ and an exterior boundary $\partial \Omega$. For the exterior boundary we consider a mix of Dirichlet and homogeneous Neumann conditions, i.e. $\partial \Omega = \partial \Omega_{\rm D} \cup \partial \Omega_{\rm N}$ with
\begin{equation}
	\left. c_{\text{p}}(\Tt{x},t) \right|_{\partial \Omega_{\text{D}}} = g(\Tt{x}), \quad \left. \Tt{n} \cdot \nabla c_{\text{p}}(\Tt{x},t) \right|_{\partial \Omega_{\text{N}}} = 0,
\end{equation}
where $\Tt{n}$ denotes the outward pointing normal vector on $\partial \Omega$.
At the phase boundaries $\Gamma$, the flux must be continuous, i.e.
\begin{equation}
	 D_{\rm lip} \nabla c_{\rm lip}(\Tt{x},t) \cdot \Tt{n} = D_{\rm cor} \nabla c_{\rm cor}(\Tt{x}, t) \cdot \Tt{n}.
\end{equation}
Along $\Gamma$, concentrations can be discontinuous. However, they are often assumed to be linked by a Nernst's equilibrium $K_{\text{cor/lip}} c_{\text{lip}}= c_{\text{cor}}$. When $K_{\text{cor/lip}}$ is constant, the model may be reformulated, e.g., in terms of a continuous concentration \cite{Rim2008,Naegel2009}. Hence this work employs the simplifying assumption $c_{\text{lip}}= c_{\text{cor}}$. Note that more complex situations with locally fluctuating or even concentration dependent coefficients also play an important role \cite{Anissimov2004,Anissimov2009,Rim2009}.

\begin{figure}[h]
	\subfigure[Time $t=\nicefrac{T}{16}$.]{\includegraphics[width=0.49\textwidth,keepaspectratio]{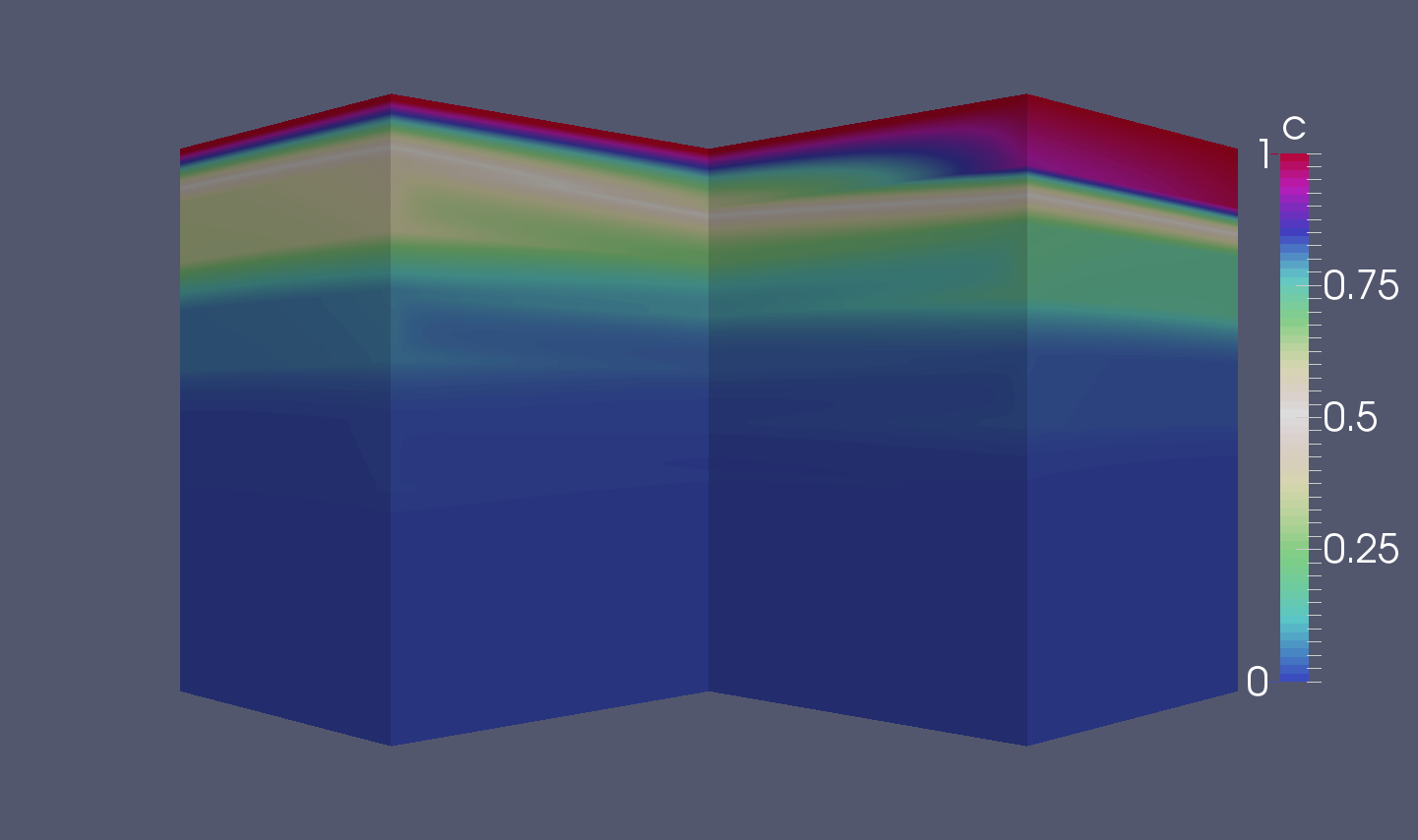}}
	\subfigure[Time $t=\nicefrac{T}{2}$.]{\includegraphics[width=0.49\textwidth,keepaspectratio]{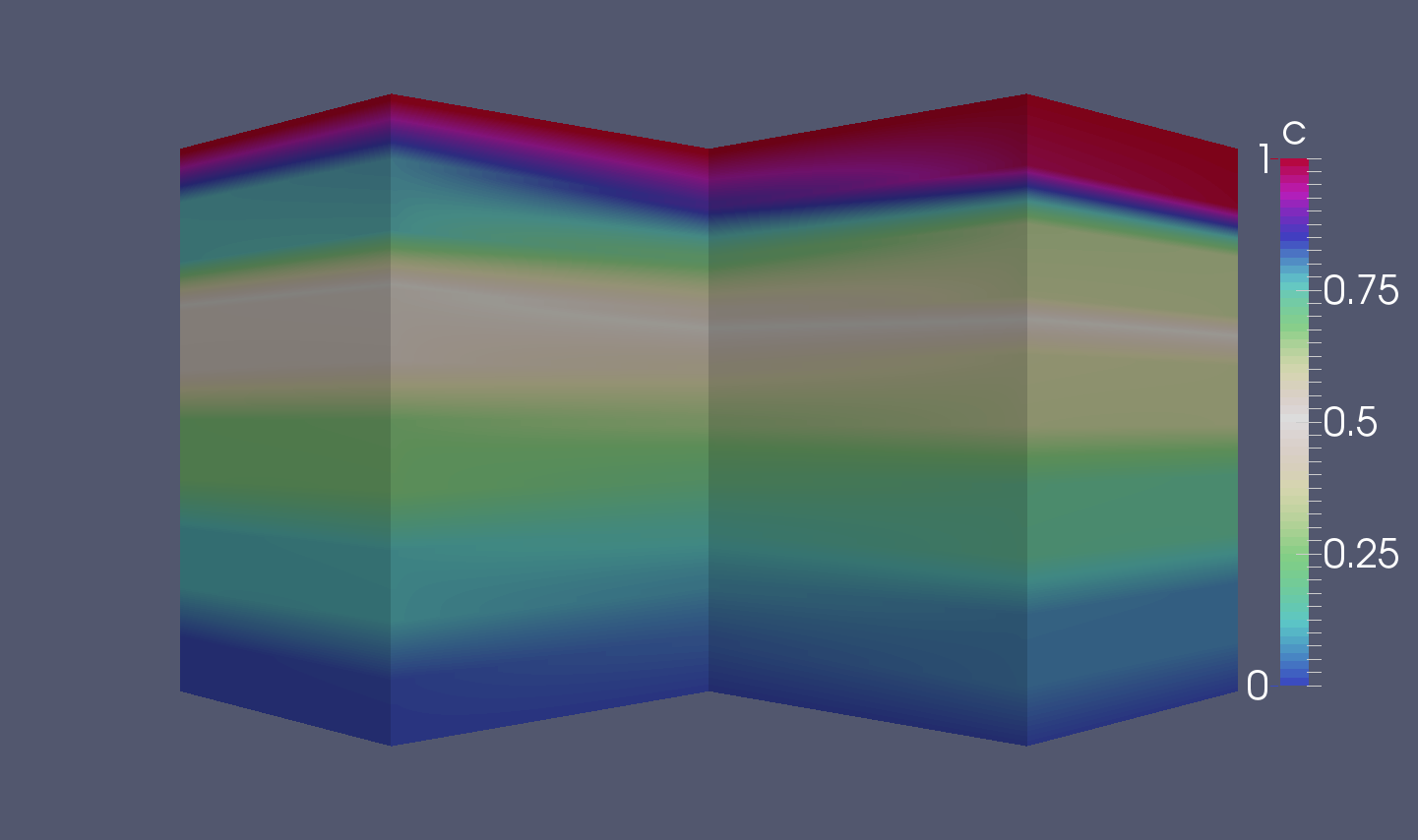}}
	\subfigure[Time $t=T$.]{\includegraphics[width=0.49\textwidth,keepaspectratio]{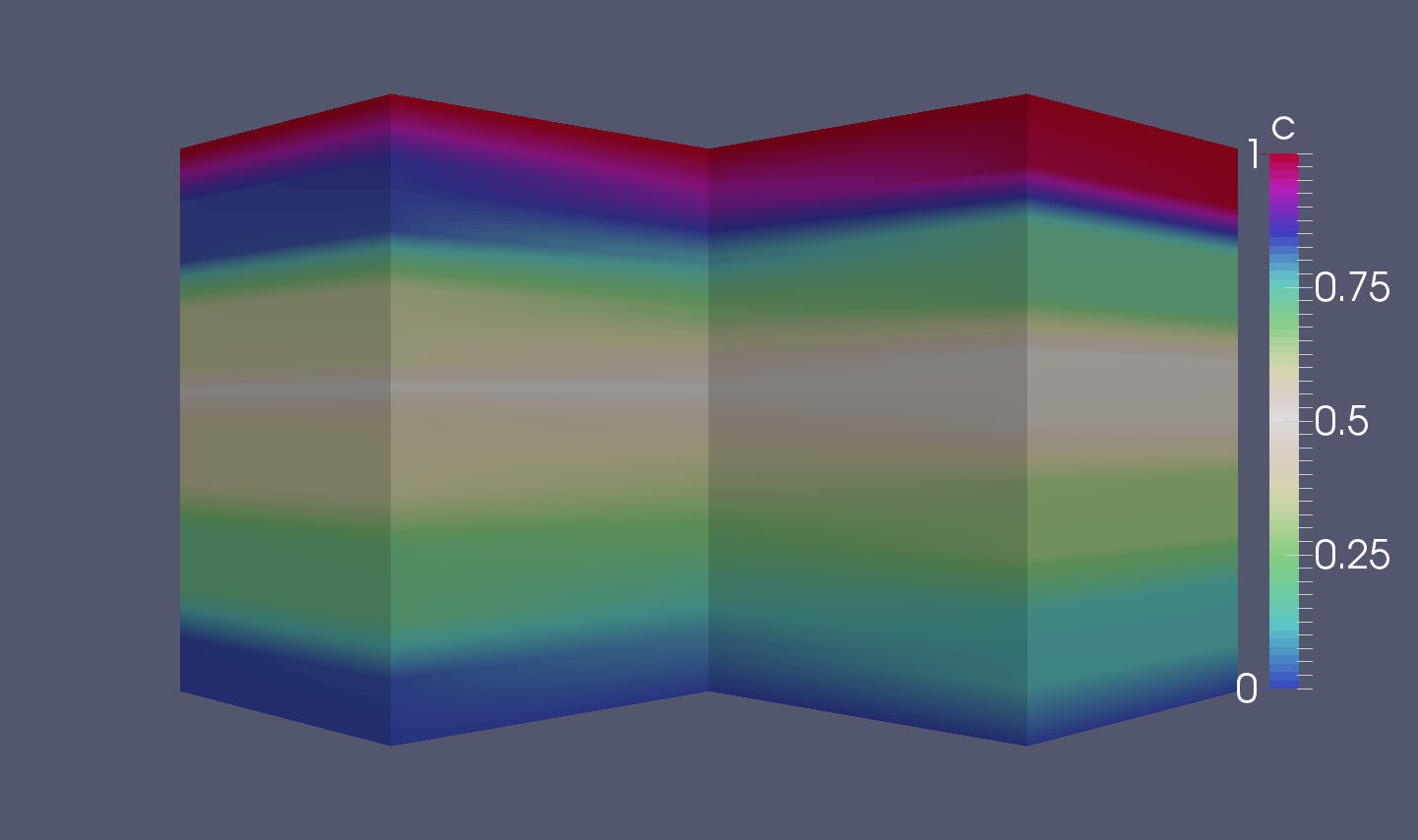}}
	\caption{Solution at $t=\nicefrac{T}{16}$ (top), $t=\nicefrac{T}{2}$ (middle), and $t=T$ (bottom) where $T$ denotes the lag-time of the problem. The block in the front has been removed to allow for a view into the interior of the computational domain.}\label{fig:solution}
\end{figure}

The geometry used in this article is a 3D brick and mortar configuration as depicted in Figure~\ref{fig:brick}. It features ten layers of corneocytes (yellow) that are embedded in a matrix of lipid bilayers (blue). It is a simplified version of more elaborate tetrakaidekahedral models with hybrid grids  presented previously~\cite{Naegel2009}; the brick and mortar model only consist of hexahedra with a reduced level of anisotropy. 
However, since jumping coefficients are present, it features some of the issues one encounters also in more complex situations.

Figure~\ref{fig:solution} shows the computed solution at three different times.
To allow for a view into the interior of the domain, a cuboid representing a quarter of the total domain has been removed in the representation.
Initially (not shown), the solution is zero on the whole domain with a Dirichlet boundary condition of $c_{\text{p}}(\Tt{x},t) = 1$ on the top side.
Then, the tracer starts to diffuse downwards in the lipid channels and much slower in comparison through the corneocytes.
In the first subfigure, diffusion has just started and filled the upper half of the domain but the concentration is still essentially zero in the lower half.
In the last subfigure, at the end of the simulation, the tracer has diffused down through the whole domain.
Concentrations continue to change with smaller changes from step to step, eventually approaching a steady state. A prospective 3D model with a fully resolved lipid-bilayer substructure, as suggested for two dimensions in \cite{Wang2006,Wang2007}, will feature a similar effect on an even smaller scale in time and space.

\subsection{Parareal}
Let the temporal domain $[0,T]$ be decomposed into $N_{\text{t}}$ subintervals $[t_j, t_{j+1}]$, $j=0, \ldots, N_{\text{t}}-1$, such that $t_0 = 0$, $t_{N_{\text{t}}-1} = T$ and
\begin{equation}
	[0,t_1] \cup \ldots \cup [t_{N_{\text{t}}-1}, T] = [0,T].
\end{equation}
Let $\mathcal{C}$ and $\mathcal{F}$ be a ``coarse'' and ``fine'' time integration method\footnote{The coarse method is often represented by $\mathcal{G}$, probably because of the French word ``gros'' for coarse.} with time step size $\Delta t$ and $\delta t \ll \Delta t$, respectively.
For the sake of simplicity assume that all subintervals have the same length and that a constant number of both $\delta t$ and $\Delta t$ steps cover a subinterval exactly.
Then, instead of serially integrating across $[0,T]$ with $\mathcal{F}$, Parareal uses the iteration
\begin{equation}
	\label{eq:Parareal}
	c^{k+1}_{n+1} = \mathcal{C}(c^{k+1}_{n}) - \mathcal{C}(c^k_n)  + \mathcal{F}(c^k_n)
\end{equation}
with $k$ as iteration index. For the first subinterval, i.e. for $[0,t_1]$, set
\begin{equation}
	c^k_0 = c_0
\end{equation}
for all $k$, where $c_0$ is the given initial value.
Note that as the iteration converges and $c^{k+1}_{n+1} - c^{k}_{n+1}$ approaches zero for all $n=0,\ldots,N_{\text{t}}-1$, the Parareal solution $c^{k+1}_{n+1}$ converges to the serial fine solution $\mathcal{F}(c_{n})$.

Formulation~\eqref{eq:Parareal} introduces concurrency because as soon as the iterates $c^k_n$ are known, the computationally expensive evaluation of the fine method can be done in parallel over all subintervals. That is, the time spent using the fine method in parallel equals the runtime of the fine method across a single subinterval instead of the full interval $[0,T]$.
However, multiple iterations are typically required and the propagation of corrections by the coarse method remains serial in time.
Speedup therefore depends on finding a cheap enough coarse integrator that still leads to convergence in a small number of iterations.

\subsection{Speedup from Parareal}
Denote by $N_{\rm c}$ the number of coarse time steps per subinterval, by $N_{\rm f}$ the number of fine time steps per subinterval and by $N_{\text{t}}$ the number of subintervals, which is assumed to be equal to the number of processors in the temporal direction.
Further, denote by $N_{\text{i}}$ the number of Parareal iterations and by $\tau^{\text{c}}$ and $\tau^{\text{f}}$ the computational runtime for a coarse or fine time step, respectively.
If one assumes that every time step takes the same amount of time, speedup from Parareal can be modeled by
\begin{equation}
        \label{eq:speedup_simple}
        S(N_{\text{t}}) \leq \cfrac{1}{ \left( 1 + \cfrac{N_{\text{i}}}{N_{\text{t}}} \right) \cfrac{N_{\rm c}}{N_{\rm f}} \cfrac{\tau^{\rm c}}{\tau^{\rm f}} + \cfrac{N_{\text{i}}}{N_{\text{t}}}}. 
\end{equation}
See e.g. \cite{Minion2010} or \cite{ArteagaEtAl2015} for a more detailed discussion of this bound.

For larger end times for the brick and mortar setup, however, this bound is too optimistic, because as the solution approaches a steady-state, the spatial solver requires fewer and fewer iterations per time step, making later time steps cheaper.
In the numerical simulations presented below, we use a final time $T$ for which the solution is still sufficiently far away from the steady-state and this effect is minimal, but we also discuss a formula valid for non-constant runtimes per time step.
Here, to simplify the notation, we omit the index range for sums, maxima and minima; it is always implied to be $n=0,\ldots,N_{\text{t}}-1$. Now, denote by $\gamma^{\text c}_{n}$ and $\gamma^{\text f}_{n}$ the cost of running the coarse and fine method across the subinterval $[t_n, t_{n+1}]$.
Then, a serial run of the fine or coarse method amounts to the duration
\begin{equation}
	\Gamma_{\text{f}} = \sum \gamma^{\text f}_n, \qquad \Gamma_{\text{c}} = \sum \gamma^{\text c}_n
\end{equation}
while a Parareal run with $N_{\text{i}}$ iterations costs
\begin{equation}
	\Gamma_{\text{P}} = \Gamma_{\text{c}} + N_{\text{i}} \gamma^{\text{x}},\qquad\gamma^{\text{x}} \equiv \max_{n} \left\{ \gamma^{\text{c}}_n + \gamma^{\text{f}}_n \right\},
\end{equation}
as the runtime for the parallel fine solve will be dominated by the subinterval with the longest simulation time.
Also, using proper pipelining, the parallel runtime of the serial coarse correction step will be governed by the most expensive subinterval for $\mathcal{C}$.
The resulting estimate for the speedup of Parareal is then
\begin{equation}
	S(N_{\text{t}})=\cfrac{\Gamma_{\text{f}}}{\Gamma_{\text{P}}} = \cfrac{1}{ \cfrac{\Gamma_{\text{c}}}{\Gamma_{\text{f}}} + N_{\text{i}} \cfrac{\gamma^{\text{x}}}{\Gamma_{\text{f}}}}
	\label{eq:speedup}
\end{equation}
This is a slight generalization of~\eqref{eq:speedup_simple} in the sense that if $\gamma_{n}^{\text{c}} = N_{\text{c}} \tau^{\text{c}}$ and $\gamma_{n}^{\text{f}} = N_{\text{f}} \tau^{\text{f}}$ are constant for all subintervals, we get
\begin{equation}
	\Gamma_{\text{c}} = N_{\text{t}} N_{\text{c}} \tau^{\text{c}}, \qquad \Gamma_{\text{f}} = N_{\text{t}} N_{\text{f}} \tau^{\text{f}},
\end{equation}
and
\begin{equation}
 	\gamma^{\text{x}} = N_{\text{c}} \tau^{\text{c}} + N_f \tau^{\text{f}}
\end{equation}
for which~\eqref{eq:speedup} simplifies to~\eqref{eq:speedup_simple}.
According to~\eqref{eq:speedup}, in the case of imbalances in the distribution of computational load across subintervals, possible speedup is limited by the subinterval with the longest runtime for both the fine and coarse method.

The optimal configuration therefore corresponds to equal computing times for all subintervals.
For explicit schemes, where the cost per time step is more or less constant, this balance is relatively easy to achieve by making sure every subinterval handles the same number of coarse and fine steps, resulting in the simple speedup model~\eqref{eq:speedup_simple}.
For implicit methods, however, cost per time step is mainly determined by the cost of the spatial solver (typically depending mainly on the number of iterations), which in turn depends on the \textit{a priori} unknown dynamics of the solution.
Therefore, naively load balancing Parareal with implicit methods based on the number of time steps alone can lead to a significant loss in efficiency.
Unfortunately, it seems that devising a proper load balancing in time for the implicit case is not straightforward and, to the best of our knowledge, has not yet been addressed in the literature. 
The easiest approach may be to use the information from the initial coarse run in Parareal to determine the size of the subintervals but this requires a non-negligible amount of implementation, might inhibit proper pipelining when requiring synchronization at some point and is thus left for future work.

\begin{figure}
	\centering
	\includegraphics{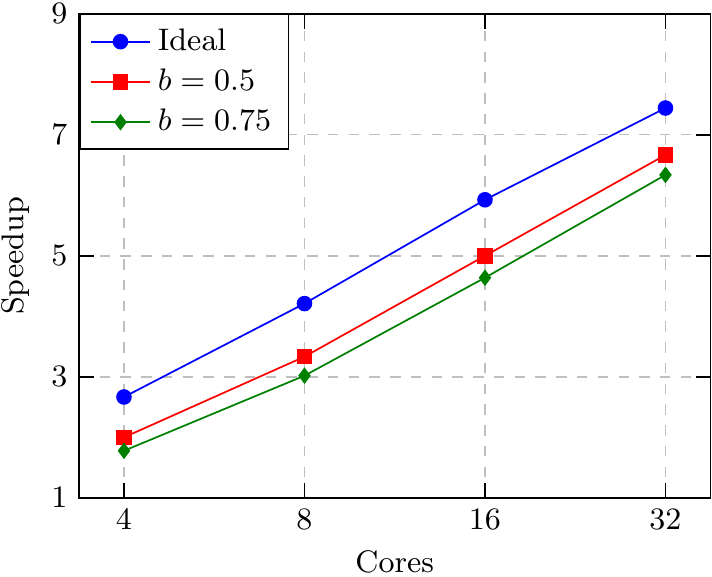}
	\caption{Projected speedup for the ideally load balanced case and a case where the second and third slice are imbalanced by a factor of $b = 0.5$ or $b=0.75$.}
	\label{fig:proj_speedup}
\end{figure}

To illustrate the effect of load imbalances in time on speedup from Parareal, Figure~\ref{fig:proj_speedup} visualizes both the projected speedup from~\eqref{eq:speedup_simple} and~\eqref{eq:speedup}: the ideal case assumes a constant coarse-to-fine ratio of
\begin{equation}
	 \cfrac{N_{\rm c}}{N_{\rm f}} = \cfrac{ \gamma^{\text{c}}_n }{ \gamma^{\text{f}}_n } = \frac{1}{10}.
\end{equation}
The imbalanced case artificially increases $\gamma^{\text{f}}_3 = (1 + b) \times 10$ and reduces $\gamma^{\text{f}}_2 = (1 - b) \times 10$ while keeping all other $\gamma^{\text{f}}_n$ and all $\gamma^{\text{c}}_n$ unchanged.
Here, $b$ is an artificial parameter modeling load imbalance between the second and third slice in the formula for projected speedup.
For $b=0$, both slices have the same load (''ideal case'') while increasing $b$ corresponds to an increasing imbalance in load: for $b=1$, the second slice no longer does any work while the third slice does twice as much work as in the ideal case.
Note that the sum $\Gamma_{\textrm{f}}$, that is the total workload, remains constant.
Clearly, the introduced imbalance has a noticeable detrimental effect on the projected speedup.

\subsection{Weak scaling}
When doubling the spatial resolution, the resulting increase in the computational cost per time step can be compensated for by a corresponding increase in the number of cores used for the spatial parallelization -- at least if both the employed method and implementation show good weak scaling.
For the spatial solver and parallelization of \texttt{UG4} applied to the benchmark used here, this has been demonstrated successfully in \cite{VogelEuroPar2015}.
However, when the number of fine time steps $N_{\text f}$ is also doubled, twice as many time steps have to be computed in order to keep the ratio $\delta t / \delta x$ constant, which leads to a doubling of time-to-solution (see also the discussion in~\cite{ArteagaEtAl2015}). 
Time parallelization can provide some mitigation by also doubling the number of subintervals and thus of cores used to parallelize along time - unless this leads to a massive increase in the number of required iterations.

However, because of the serial coarse correction, just as Parareal cannot achieve ideal strong scaling, it can also not provide $100\%$ efficiency in weak scaling.
To see this, let $h$ denote a parameter governing accuracy of the discretization. 
Estimated runtime for a Parareal run on $N_{\text{t}}$ cores (in time) is then
\begin{equation}
	R_{2h}(N_{\text{t}}) = N_{\text{t}} N_{\text{c}} \tau^{\text{c}}_{2h} + N_{\text{i}} \left( N_{\text{c}} \tau^{\text{c}}_{2h}+ N_{\text{f}} \tau^{\text{f}}_{2h} \right).
\end{equation}
Doubling spatial resolution, i.e. using $h$ instead of $2h$ and performing twice as many fine and coarse steps on twice as many subintervals (but keeping the number of coarse and fine steps per subinterval constant) gives
\begin{equation}
	R_h\left( 2 N_{\text{t}}  \right) = 2 N_{\text{t}} N_{\text{c}} \tau^{\text{c}}_h + N_{\text{i}} \left( N_{\text{c}} \tau^{\text{c}}_h+ N_{\text{f}} \tau^{\text{f}}_h \right)
\end{equation}
assuming the number of iterations does not change.
In the case of perfect weak scaling in space, by increasing the number of spatial cores the runtime per step can be kept constant, that is $\tau^{\text{c}}_{2h} = \tau^{\text{c}}_{h}$ and $\tau^{\text{f}}_{2h} = \tau^{\text{f}}_{h}$.
Under this assumption, the projected weak scaling efficiency of the space-time parallelization is
\begin{equation}
	\frac{R_{2h}(N_{\text{t}})}{R_{h}(2 N_{\text{t}})} = \frac{ N_{\text{t}} \sigma + N_{\text{i}} \left( 1 + \sigma \right) }{ 2 N_{\text{t}} \sigma + N_{\text{i}} \left( 1 + \sigma \right)} < 1
\end{equation}
with $\sigma := N_{\text{c}} \tau^{\text{c}}_{h} / N_{\text{f}} \tau^{\text{f}}_{h} > 0$.
Therefore, while weak scaling can never be perfect, a cheap enough coarse integrator ($\sigma \ll 1$) should still allow for good weak scaling -- as long as the underlying spatial solver shows good weak scaling and the number of iterations is not affected by the increasing number of subintervals.

\subsection{Implementation}
For Parareal we use the C{\tt++} library \texttt{Lib4PrM} that uses MPI for the necessary communication of volume data in time.
A straightforward approach to implementing~\eqref{eq:Parareal} is sketched as pseudo code e.g. in~\cite{ArteagaEtAl2015}.
Here, however, we use a somewhat more elaborate implementation that is based on the following observation. After the first iteration on the first subinterval $[0,t_1]$, the coarse terms in the Parareal iteration~\eqref{eq:Parareal} cancel out, resulting in
\begin{equation}
	c^{k+1}_1 = \mathcal{F}\left( c_0 \right)
\end{equation}
for $k \geq 1$. That is, after one iteration, the first subinterval is guaranteed to have converged and the processors responsible for the first subinterval can ``retire''. 
After the second iteration, by the same argument, this will then be true for the processors handling the second subinterval and so on. 
After $k$ iterations, the time-parallel fine method is guaranteed to have converged on the first $k$ subintervals and all processors with an MPI rank (in time) smaller or equal to $k$ could in principle be used otherwise.
Put differently, Parareal converges always at least at a rate of one subinterval per iteration and when $k=n$ the Parareal method is guaranteed to have converged at $t_k\leq t_n$. While leaving processors idle according to this implementation does not affect runtime negatively, it has the potential to reduce the energy cost of a simulation, particularly in combination with ``dynamic voltage and frequency scaling''~\cite{Dick2015}.
This will be studied in a future work~\cite{kreienbuehl2015a}.
Also, if not enough processors are available to cover the whole interval $[0,T]$ by subintervals of a given size, converged processors could pick up subintervals at the end in a caterpillar-like way.
Such even more involved implementations are left for future studies, though.
Finally, in production runs one could also use some tolerance e.g. for the updates between iterations or a proper residual to decide when the solution at the end of a subinterval is converged~\cite{Ruprecht2014_GAMM}.

\subsection{Spatio-temporal discretization and solvers}
Discretization in space and time is provided by the software package \texttt{UG4} \cite{VogelUG2013}. We employ a plain vanilla vertex centered finite volume scheme in space that is combined with an implicit Euler scheme in time. For each time step this gives rise to a large linear system of equations, where the number of degrees of freedom corresponds to the number of vertices of the grid. The solver is a multi-grid method with three steps of damped ($\omega=0.6$) Jacobi relaxation used for pre- and post-smoothing. More details are provided in \cite{VogelEuroPar2015}. The coarse grid problem with 7`581 degrees of freedom was solved using sequential SuperLU \cite{superlu_ug99,superlu99}.

\section{Numerical results}
We report results from solving the brick and mortar problem described in \S\ref{sec:methods} with the simulation framework described above.
For both $\mathcal{C}$ and $\mathcal{F}$ we use an implicit Euler method with the time step size $\Delta t$ for $\mathcal{C}$ being significantly larger than the time step size $\delta t$ for $\mathcal{F}$.

All runs are performed on the Cray XC40 Piz Dora supercomputer at the Swiss National Supercomputing Centre (CSCS) in Lugano, Switzerland. This supercomputer is equipped with 1`256 compute nodes, each of which consists of two 12-core Intel Xeon E5-2690v3 CPUs, making for a total of 30`144 compute cores.\footnote{\url{http://user.cscs.ch/computing_systems/piz_dora/}} Its peak performance is 1.254 PFlops, placing it at position $56$ in the Top500 November, 2014 list.\footnote{\url{http://www.top500.org/list/2014/11}}
As compiler we used version 4.9.2 of the the GNU compiler collection\footnote{\url{https://gcc.gnu.org}} and, in the following, report runtimes of simulations without I/O operations.

\subsection{Convergence of Parareal}
Generally speaking, convergence of Parareal is affected by a number of parameters: The time step sizes and methods used for $\mathcal{C}$ and $\mathcal{F}$, the number of concurrently computed subintervals, the discretization used for the spatial derivatives, and the dynamics of the problem to be solved.
To measure convergence, below the relative defect $d^k_n$ between Parareal after $k$ iterations and the fine solution run in serial is reported, i.e.
\begin{equation}
	d^k_n = \cfrac{ \left\| c^k_n - c_n \right\|_2}{ \left\| c_n \right\|_2}, \ n=0, \ldots, N_{\text{t}}-1
\end{equation}
with
\begin{equation}
	c_n = \mathcal{F}(c_{n-1}), \ n=1,\ldots,N_{\text{t}}-1.
\end{equation}
In order to avoid distortions through I/O times, only the final solution values are written out and the defect $d^k_{N_{\text{t}}-1}$ is reported except for in section \S\ref{subsec:error_over_time}.
There, the defect is analyzed not only as a function of the number of iterations but also time.

\begin{figure}[t]
    \centering
    \includegraphics{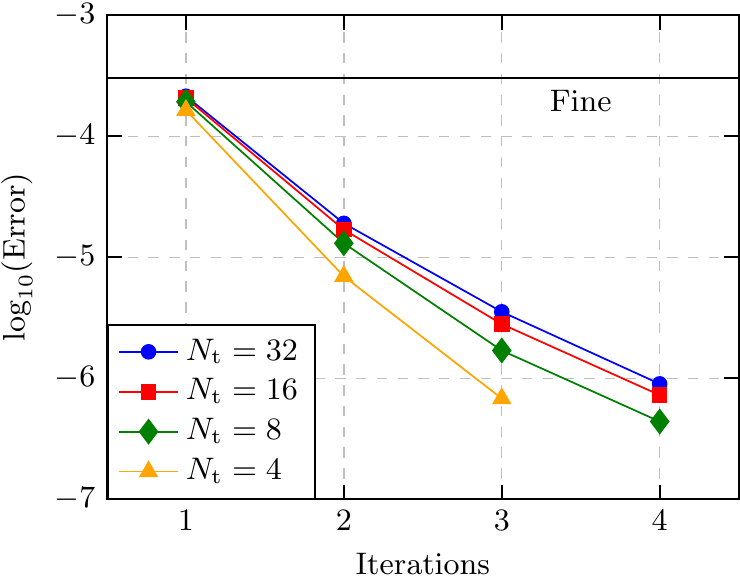}
    \caption{The defect $d^k_{N_{\text{t}}-1}$ between the Parareal and the serial fine solution at $t=T$ versus the number of Parareal iterations for different numbers of subintervals $N_{\text t}$ is illustrated. The discretization error of the serial fine method is indicated by the black horizontal line.}
    \label{fig:convergence}
\end{figure}

Figure~\ref{fig:convergence} shows the defect $d^k_{N_{\text{t}}-1}$ versus the number of iterations $k$.
In addition, the estimated discretization error of the fine method resulting from a comparison against a run of $\mathcal{F}$ with a time step four times smaller than $\delta t$ is shown.
Parareal converges rapidly in all configurations.
As can be expected, computing more subintervals in parallel slows down convergence somewhat. 
Here, for all $N_{\text t}\in\{4,8,16,32\}$, one iteration suffices to reduce the defect below the discretization error of the fine method.
\textit{This confirms the usability of Parareal also for complex diffusion problems with anisotropic geometries and large jumps in the coefficients.}
Particularly the relatively mild reduction of convergence speed as $N_{\text{t}}$ is increased illustrates the potential for using larger numbers of cores to parallelize in time for this kind of problem, should a sufficiently large machine be available.

\subsection{Effect of spatially varying coefficients}
\begin{figure}[t]
    \centering
    \includegraphics{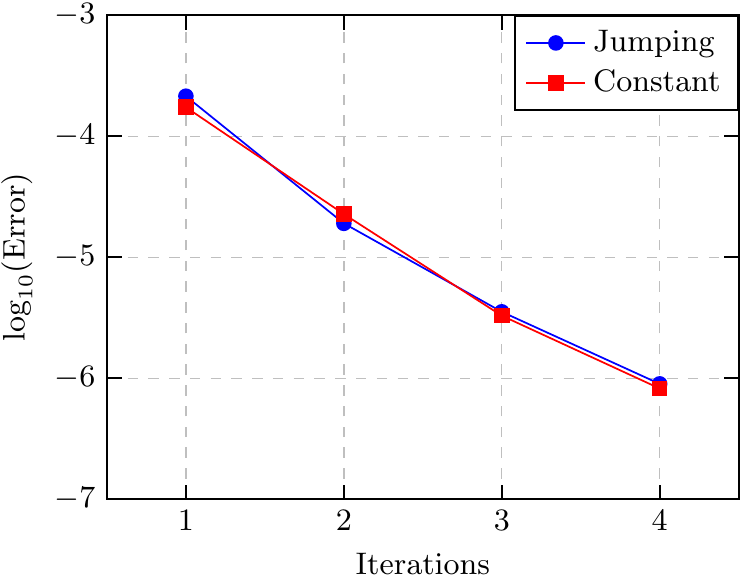}
    \caption{The defect $d^k_{N_{\text{t}-1}}$ at $t=T$ between the Parareal $N_{\text t} = 32$ solution and the fine serial solution versus the number of Parareal iterations for the brick and mortar problem (red) and a reference configuration with constant diffusion coefficients (blue) is shown.}\label{fig:const_coeff}
\end{figure}

In the brick and mortar problem, the diffusion coefficients jump between $D_{\text{lip}}=1$ in the lipid channels and $D_{\text{cor}} = 10^{-3}$ in the corneocytes.
To assess the impact these jumps have on the convergence of Parareal, Figure~\ref{fig:const_coeff} gives a comparison of the defect for the brick and mortar problem (red) and a reference configuration with $D_{\text{lip}} = D_{\text{cor}} = 10^{-3}$ throughout the whole domain.
For the setup studied here, in line with the findings for 2D problems in~\cite{RuprechtEtAl2015_DDM}, the jump in coefficients has almost no effect on how Parareal convergence.
Experiments not documented here suggest that a larger $T$ (that is, a final configuration closer to the steady state) can lead to a larger detrimental effect of coefficient jumps: However, even there this only resulted in a small number of additional iterations required for convergence.

\subsection{Error over time}\label{subsec:error_over_time}
\begin{figure}[t]
    \centering
    \includegraphics{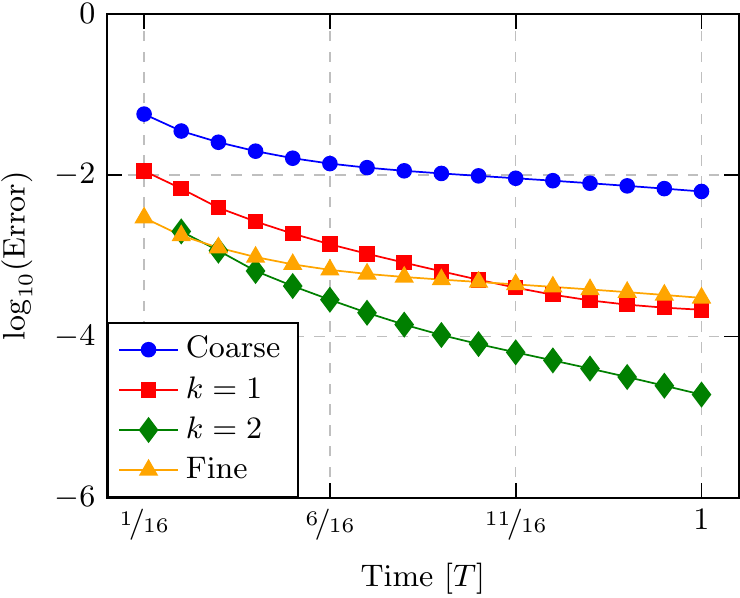}
    \caption{The discretization error over time for the fine and coarse solution, and the defect of Parareal with $N_{\text t} = 32$ subintervals after $k=1$ and $k=2$ iterations is illustrated.}
    \label{fig:error_over_time}
\end{figure}

So far only the defect at the end of the simulation has been reported. 
In contrast, Figure~\ref{fig:error_over_time} shows both the defect $d^k_n$ of Parareal for $k=1$ (red) and $k=2$ (green) as well as the estimated discretization error of the coarse (blue) and fine (yellow) integrator.
The figure shows the defect after every second subinterval for Parareal using $N_{\text{t}} = 32$ .

Already after one iteration, the solution at $T=1$ provided by Parareal has the same quality as when running the fine method serially.
Therefore, speedup is reported below using $k=1$.
However, one iteration is not sufficient to reduce the defect of the whole transient to the discretization error: here, two iterations would be required after which the green line (Parareal) is completely below the yellow line (fine integrator).

\subsection{Scaling of Parareal}\label{subsec:scaling}
\begin{figure}[t]
    \centering
    \includegraphics{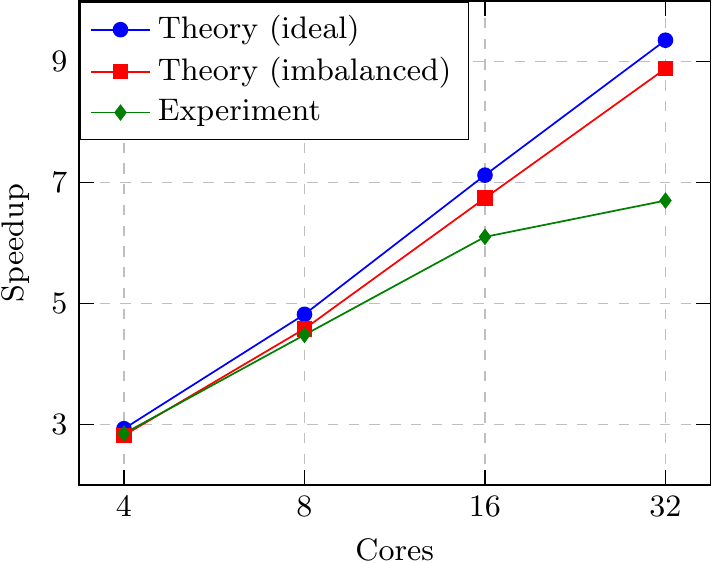}
    \caption{Depicted is the speedup of the time parallelization with Parareal.}\label{fig:speedup}
\end{figure}

Figure~\ref{fig:speedup} shows the speedup from Parareal compared to running $\mathcal{F}$ in serial with the number of iterations chosen such that the defect of Parareal at $T=1$ is below the estimated discretization error of $\mathcal{F}$ (for $N_{\text{t}}\in\{4,8,16,32\}$ this means one iteration).
The projected speedups for ideal load balancing according to~\eqref{eq:speedup_simple} are marked by blue circles while the projected speedups according to~\eqref{eq:speedup}, including differences in runtime between subintervals, are shown as red squares.
Here, runtimes per subinterval $\gamma_{n}^{\text{c}}$ and $\gamma_{n}^{\text{f}}$ are measured experimentally from serial runs of $\mathcal{C}$ and $\mathcal{F}$.
Measured speedups are shown as green diamonds.
Up to sixteen subintervals, speedup follows the projected value reasonably well, but for $32$ subintervals noticeable drop-off is observed -- in small parts, this is due to imperfect load balancing as indicated by the difference between the red and blue curve.
The major part, however, is overhead from communication and other factors, which are not incorporated in the speedup model.

\subsection{Spatio-temporal weak scaling}
\begin{figure}[t]
    \centering
    \includegraphics{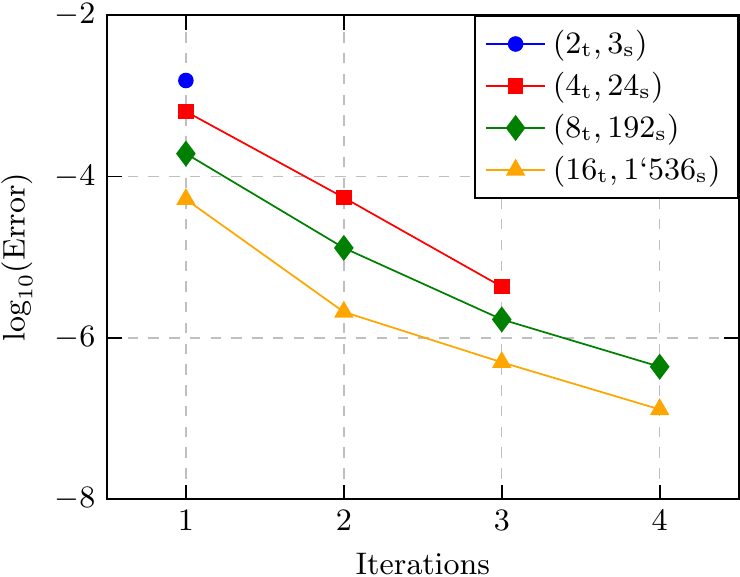}
    \caption{The convergence of Parareal for four different configurations is shown. The ratios $\delta t/\delta x$ and $\Delta t/\delta x$ are kept constant and the number of degrees-of-freedom per core as well as the number of time steps per subinterval are kept constant, too. Thus, e.g. in the run with $8$ cores in time and $192$ cores in space the spatio-temporal resolution is twice as good as in the run with $4$ cores in time and $24$ cores in space. Note that because the problem is in 3D, doubling the spatial resolution requires eight times more cores.}
    \label{fig:weak_scaling}
\end{figure}

%
%
\begin{table*}[th]
	\centering
	\begin{tabular}{lrrrrrrrrrrr}
		\toprule
		Run & $\Delta t^{-1}$ & $\delta t^{-1}$ & \# Elements & $P_{\text{space}}$ & $N_{\text{t}}$ & $P_{\text{total}}$ & $e_{\text{fine}}$ & $d^1_{N_{\text{t}}}$ & $R$\ {[}s{]} & Factor \\
		\midrule
		$(2_{\text{t}},3_{\text{s}})$ & 8 & 128 & 277`440 & 3 & 2 & 6 & $10^{-2.9}$ & $10^{-2.8}$ & 132.79 & -- \\
		$(4_{\text{t}},24_{\text{s}})$ & 16 & 256 & 2`219`520 & 24 & 4 & 96 & $10^{-3.2}$ & $10^{-3.2}$ & 239.34 & 1.80 \\
		$(8_{\text{t}},192_{\text{s}})$ & 32 & 512 & 17`756`160 & 192 & 8 & 1`536 & $10^{-3.5}$ & $10^{-3.7}$ & 316.83 & 1.32 \\
 		$(16_{\text{t}},1`536_{\text{s}})$ & 64 & 1`024 & 142`049`280 & 1`536 & 16 & 24`576 & $10^{-3.8}$ & $10^{-4.3}$ & 508.70 & 1.61 \\
		\bottomrule
	\end{tabular}
	\caption{Configuration of the runs shown in Figure~\ref{fig:weak_scaling}. Both the number of coarse and fine time steps per core in time and the number of elements per core in space are kept constant in all runs. Here, $e_{\text{fine}}$ indicates the estimated discretization error of the fine integrator and $d_{N_{\text{t}}}^{1}$ is the defect after one iteration. $R$ indicates runtime in seconds. Note that $N_{\text{t}}$ is the number of subintervals and equal to the number of cores in time. The last column gives the factor between runtimes: ideal space-time weak scaling corresponds to a factor of $1.0$, ideal spatial weak scaling with no time parallelization corresponds to a factor of two while no weak scaling at all would lead to a factor of $8 \times 2 = 16$ because the simulations use a 3D spatial discretization.}
	\label{tab:weak_scaling}
\end{table*}
Figure~\ref{fig:weak_scaling} shows convergence of Parareal runs for four different setups with increasing spatial and temporal resolution but keeping both the number of elements in space and time steps per core constant. 
The first one (blue) uses a time step size of $\Delta t = \nicefrac{1}{8}$ and $\delta t = \nicefrac{1}{128}$ in units of $T$.
The second one (red) on the other hand uses half the coarse and fine time step and half the spatial mesh width so that $\delta t/\delta x$ and $\Delta t/\delta x$ are the same in both runs.
It also uses twice as many cores in time and eight times more cores in space, so that both the number of elements per core and the number of time steps per core remain constant, too.
The green and yellow line then correspond to runs that again double spatial and temporal resolution.
Higher spatio-temporal resolution leads to smaller defects for Parareal, while the rates of convergence (i.e. the slopes of the lines) remain roughly the same.

Exact configurations are shown in Table~\ref{tab:weak_scaling}: note how $\Delta t^{-1} / N_{\text{t}}$ and $\delta t^{-1} / N_{\text{t}}$ as well as $\text{\# Elements} / P_{\text{space}}$ are constant in all configurations.
Also, each refinement step halves the estimated fine discretization error, which matches the behavior expected for the employed first-order discretization.
The number of Parareal iterations required for convergence to the accuracy of $\mathcal{F}$ stays constant: every configuration is converged after one iteration.
Runtimes are increasing as the problem size grows, so space-time weak scaling is not optimal.
Partly, this is because of the overhead from the coarse method, see the discussion above, partly because of less than optimal weak scaling of the spatial solver.
Nevertheless, Parareal helps to mitigate some of the increase in runtime from increasing the spatio-temporal problem size.

\section{Conclusions}
Computational modeling of skin permeation is of interest for different applications. 
However, the full problem requires resolution of a vast range of scales, leading to enormous computational requirements.
Massively parallel computers are needed but these require suitable parallel numerical methods to be used efficiently.

This article investigates the applicability and performance of the time-parallel Parareal integrator to a relevant precursory problem of skin transport, namely a 3D brick and mortar configuration.
For this, the C{\tt++} Parareal library \texttt{Lib4PrM} is integrated as a plug-in into the simulation environment \texttt{UG4} using implicit integrators and a geometric multi-grid as spatial solver.
While the brick and mortar problem does not yet feature the same geometric level of detail as the skin transport problem, it already has jumps in the diffusion coefficients of several orders of magnitude on a highly an\-is\-otr\-op\-ic domain.
The article is an extension of a previous study of a 2D problem on a domain with a much simpler structure~\cite{RuprechtEtAl2015_DDM}.

Performance of the space-time parallel solver is studied in several numerical experiments.
It is confirmed that Parareal still converges quickly for the brick and mortar case.
Moreover, strong and weak scaling as well as implications for the ``trap of weak scaling'' are illustrated and discussed.

As the solution of the brick and mortar problem approaches a steady state in time, initial guesses for the geometric multi-grid become more accurate if time steps of constant length are used.
This leads to faster convergence of the geometric multigrid, which in turn induces an imbalance in workload between the different processors in time.
For the chosen setup, this effect is small but by deriving a simple theoretical model, imbalances in time are shown to have a potentially significant effect on parallel efficiency.
For Parareal with implicit integrators applied to complex PDEs, the resulting load imbalance is an important issue that has to be addressed.

A number of possible directions for future research emerge from the experiments presented here.
So far, coarsening in Parareal was done only in time by using a larger time step. 
Better results can be expected if the spatial discretization is coarsened simultaneously.
This requires a closer integration of Parareal with the spatial multi-grid solver, in order to provide interpolation and restriction routines.
Also, this approach can be taken further by interweaving iterations of the time-parallel method with iterations of the spatial solver, as discussed e.g. for Parareal in~\cite{Mula2014} or for PFASST in~\cite{MinionEtAl2015}.
Another important issue that is also connected to load balancing is spatial and temporal adaptivity. While both can in principle be used in Parareal, they greatly complicate the load balancing problem.
Finally, as energy consumption is becoming a more and more important issue in high-performance computing, a thorough benchmarking in terms of energy-to-solution is also an important direction for future work.

\bibliographystyle{spmpsci}      
\bibliography{Pint,Skin,bibtex}   

\end{document}